\documentclass{Phbauth}
\usepackage{graphicx}

\begin{document}

\begin{frontmatter}

\title{Magnetic properties of Gd$_{1-x}$Pr$_x$Ba$_2$Cu$_3$O$_{7-y}$
single crystals 
.}

\author[address2]{V. N. Narozhnyi $\rm ^{a,}$\thanksref{thank1}},
\author[address1]{D. Eckert},
\author[address1]{G. Fuchs},
\author[address1]{K. Nenkov\thanksref{thank2}},
\author[address3]{T. G. Uvarova},
\author[address1]{K.-H. M\"uller}

\address[address1]{Institut f\"ur Festk\"orper- und Werkstofforschung 
Dresden e.V., 270016, D-01171 Dresden, Germany}

\address[address2]{Institute for High Pressure Physics, Russian 
Academy of Sciences, Troitsk, Moscow Reg., 142092, Russia}

\address[address3]{Institute of Crystallography, Russian 
Academy of Sciences, Leninski pr. 59, Moscow, 117333, Russia}

\thanks[thank1]{Corresponding author. Present address: Institute for
High
Pressure Physics, Russian Academy of Sci., Troitsk, Moscow Reg.,
142092,
Russia. Fax:+7-(095)-334-00-12. E-mail: narozh@ns.hppi.troitsk.ru}

\thanks[thank2]{On leave from Int. Lab. HMFLT, Wroclaw, 
Poland and Institute for Solid State Physics BAS, Sofia, Bulgaria.}

\begin{abstract}
Magnetic properties were studied for the high quality Al-free
orthorhombic Gd$_{1-x}$Pr$_x$Ba$_2$Cu$_3$O$_{7-y}$ single crystals
($0\leq x \leq 1.0$) grown by the flux method. An indication on the
interaction between the Pr and Cu(2) magnetic sublattices was found
for Pr123. Different sign of magnetic anisotropy was established for
the Pr and Gd ions at low $T$. It was also shown that
superconductivity reported by Zou $\mathit et~al.$ [Phys. Rev. Lett.
80, 1074 (1998)] for Pr123 single crystals grown by TSZF method seems
to be connected with partial substitution of Ba for the Pr-sites.
\end{abstract}

\end{frontmatter}


$\rm GdBa_2Cu_3O_{7-y}$ (Gd123) has typical behavior for fully doped
orthorhombic $R\rm Ba_2Cu_3O_{7-y}$ ($R$=Y, rare earth) high-$T_c$
cuprates ($T_c\approx 90~$K, $T_N\approx 2.2~$K), whereas Pr123 with
$T_ N \approx 17~$K is the anomalous member among $R$123
\cite{Radousky}.

In this work we have studied the magnetic properties of the mixed $\rm
Gd_{1-x}Pr_xBa_2Cu_3O_{7-y}$ [(Gd-Pr)123] and (Y-Pr)123 compounds.
High quality Al-free single crystals with ($0\leq x \leq 1.0$) were
grown in Pt crucibles by the flux method \cite{Nar_JMMM96}. Atomic
absorption spectroscopy has shown that the Pt contamination does not
exceed $3~10^{-3}$~at.~\%. 

\begin{figure}[btp]
\begin{center}\leavevmode
\includegraphics[width=0.8\linewidth]{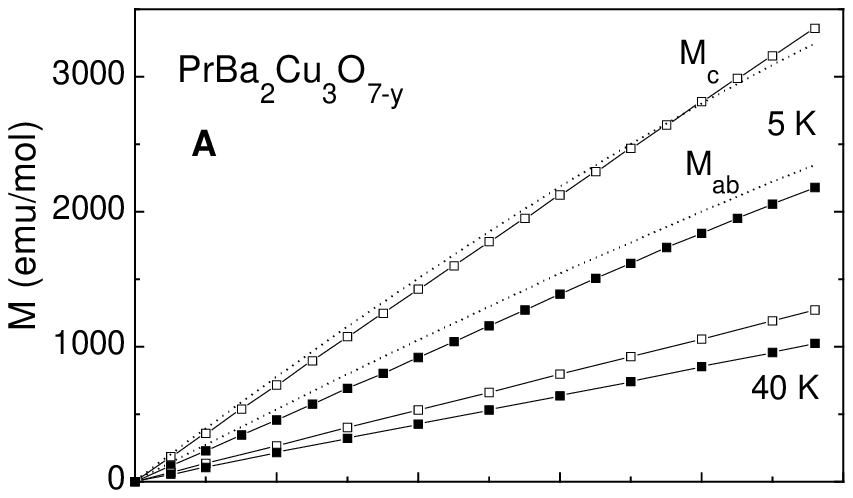}

\vspace{0.6cm}

\includegraphics[width=0.8\linewidth]{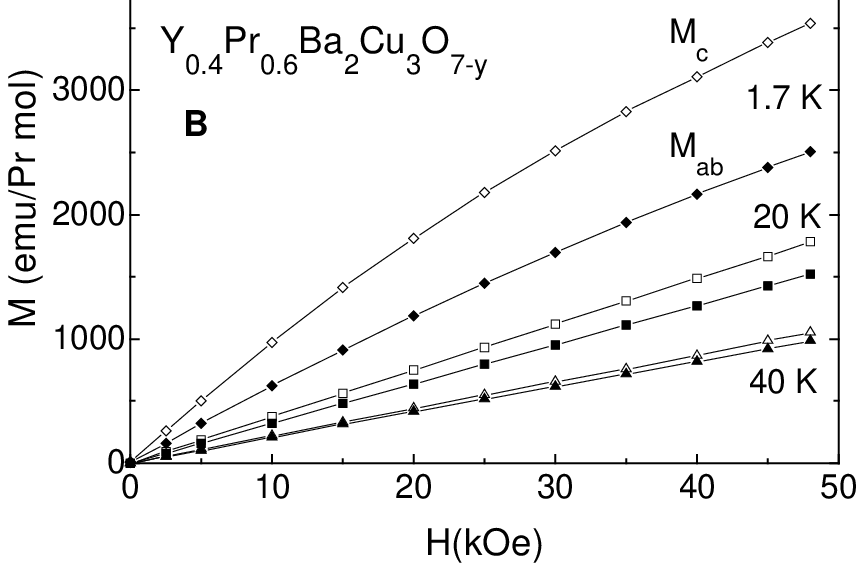}
\protect\caption{ $M$ vs $H$ for the two directions of $H$ for a
Pr-123
(A) and $\rm (Y_{0.4}Pr_{0.6})123$ (B) single crystals. Dotted lines 
show $M$ vs. $H$ (in emu/mol~Pr) for a $\rm (Y_{0.4}Pr_{0.6})123$ at
$T=5~$K.}
\label{fig_YPr}
\end{center}
\end{figure}

\begin{figure}[btp]
\begin{center}\leavevmode
\includegraphics[width=0.8\linewidth]{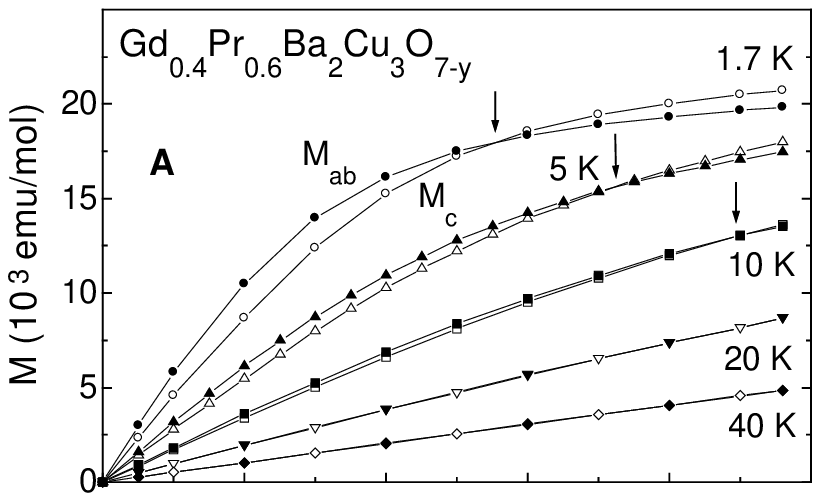}

\vspace{0.6cm}

\includegraphics[width=0.8\linewidth]{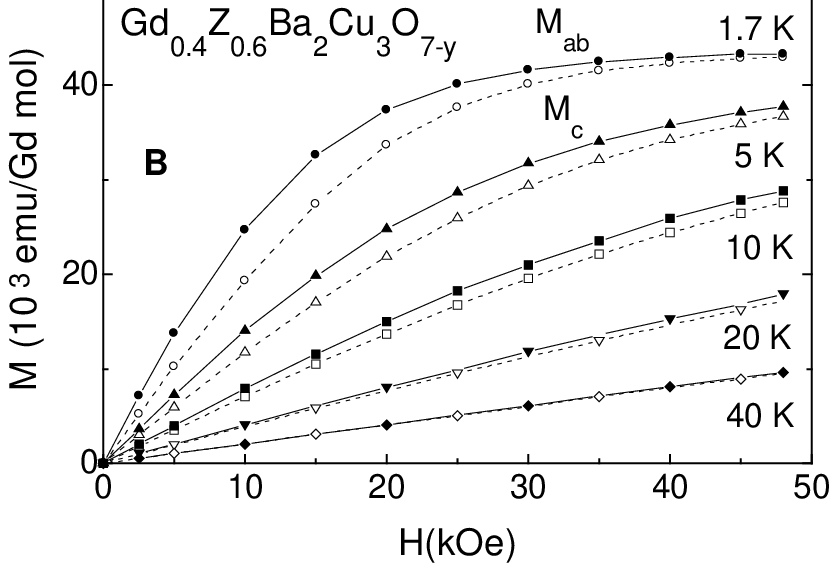}
\protect\caption{$M$ vs $H$ for the two directions of $H$ for a $\rm
(Gd_{0.4}Pr_{0.6})123$ 
(A) and imaginary $\rm (Z_{0.4}Pr_{0.6})123$ (B) single crystals. See
text for details.}
\label{fig_GdPr}
\end{center}
\end{figure}

Magnetization curves $M_c(H)$ ($H||c$) and $M_{ab}(H)$ [$H||(ab)$] for
($\rm Y_{0.4}Pr_{0.6})123$ crystal are rather similar to the obtained
for pure Pr123 crystals, see Fig.\ \ref{fig_YPr}. The sign of magnetic
anisotropy is different for ($\rm Gd_{0.4}Pr_{0.6})123$ at low $H$ and
low $T$ ($M_{ab}>M_c$), see Fig.\ \ref{fig_GdPr}A. A crossover to
$M_{ab}<M_c$ can be seen in high $H$ at $T \leq 10~$K (marked by
arrows in the Fig.\ \ref{fig_GdPr}A. To obtain the contribution to
$M(H)$ from Gd ions the results for ($\rm Y_{0.4}Pr_{0.6})123$ were
substracted from the corresponding curves for ($\rm
Gd_{0.4}Pr_{0.6})123$, see Fig.\ \ref{fig_GdPr}B, where the results
are market as belonging to imaginary ($\rm Gd_{0.4}Z_{0.6})123$
compound with Z marking empty spaces on Gd sites. It is clear that
$M_{ab}>M_c$ for Gd-sublattice and the value of magnetic anisotropy
decreases gradually with increase of $T$ or $H$. At $T \geq 40~$K the
anisotropy is close to zero as could be expected for Gd$^{3+}$ ions
having only spin component of magnetic moment. At $T=1.7~$K in
$H=48~$kOe the magnetization is practically isotropic, see Fig.\
\ref{fig_GdPr}B. Therefore the observed crossover of magnetic
anisotropy for ($\rm Gd_{0.4}Pr_{0.6})123$ is connected with different
signs of anisotropy for the Pr- and Gd-sublattice in this compound and
with the different character of $M(H)$ dependencies for them.

It was found, that the kink in magnetic susceptibility $\chi_{ab}(T)$
connected with AFM ordering of Pr disappears after field cooling (FC)
in $H\parallel ab$-plane, whereas the kink in $\chi_c(T)$ remains
unchanged after FC in $H\parallel c$-axis. Possible explanation is
connected with coupling of Pr and Cu sublattices because Pr ordering
is accompanied by a reordering of Cu moments below $T_N$
\cite{Boothroyd_PRL97}. The theory of exchange-frustrated
antiferromagnets with two spin subsystems interacting only by the
anisotropic pseudodipole interaction has been recently proposed by
S.V. Maleev \cite{Maleev_JETPL}.

It was generally accepted that Pr-123 is the only nonsuperconducting
compound in R123 row, but recently Zou $\mathit et~al.$
\cite{Zou_PRL98} reported the observation of bulk superconductivity
for Pr-123 grown by the traveling-solvent floating-zone (TSZF) method.
Our analysis of Zou's data \cite{Zou_PRL98} has shown \cite{Nar_Comm}
that the Curie constant $C$ for his crystal is about two times smaller
than obtained for our Pr123 crystals \cite{Nar_PhC99} or reported by
other groups \cite{Uma_JPh98}. This suggests that Pr occupies only
about one half of the $R$ sites. The other half is occupied most
probably by the nonmagnetic Ba.

This work was supported by RFBR grant 96-02-00046G and DFG grant 
MU1015/4-1.

\end{document}